\documentclass{aa}

\usepackage{pdflscape}
\usepackage{graphicx}
\usepackage{txfonts}

\begin{document} 

   \title{IR nebulae around bright massive stars \\ as indicators for binary interactions}

   \author{J. Bodensteiner\inst{1,2,3}, D. Baade\inst{4}, J. Greiner\inst{3}, and N. Langer\inst{5}}

   \institute{Instituut voor Sterrenkunde, KU Leuven, Celestijnenlaan 200D, Bus 2401, 3001 Leuven, Belgium\\
              \email{julia.bodensteiner@kuleuven.be}
        \and
             Technische Universität München, James-Franck-Straße 1, 85748 Garching, Germany
         \and
             Max-Planck-Institut für extraterrestrische Physik, Gießenbachstraße, 85748 Garching, Germany
         \and
             European Southern Observatory, Karl-Schwarzschild-Straße 2, 85748 Garching, Germany
         \and
                 Argelander-Institut für Astronomie der Universität Bonn, Auf dem Hügel 71, 53121 Bonn, Germany
        } 
   \date{Received Month xx, xxxx; accepted Month xx, xxxx}

 
  \abstract
   {Recent studies show that more than 70\% of massive stars do not evolve
    as effectively single stars, but as members of interacting binary systems. The
    evolution of these stars is thus strongly altered compared to similar but
    isolated objects.}
   {We investigate the occurrence of parsec-scale mid-infrared nebulae around
   early-type stars. If they exist over a wide range of stellar properties, one possible
   overarching explanation is non-conservative mass transfer in binary interactions,
   or stellar mergers.}
   {For $\sim3850$ stars (all OBA stars in the Bright Star Catalogue [BSC], Be stars, 
   BeXRBs, and Be+sdO systems), we visually inspect WISE 22\,$\mu$m images. Based on
   nebular shape and relative position, we distinguish five categories: offset bow shocks
   structurally aligned with the stellar space velocity, unaligned offset bow shocks,
   and centered, unresolved, and not classified nebulae.}
   {In the BSC, we find that 28\%, 13\%, and 0.4\% of all O, B, and A stars, respectively, 
   possess associated infrared (IR) nebulae. Additionally, 34/234 Be stars, 4/72 BeXRBs, and 3/17 
   Be+sdO systems are associated with IR nebulae. }
   {Aligned or unaligned bow shocks result from high relative velocities between star
   and interstellar medium (ISM) that are dominated by the star or the ISM, respectively.  About 
   13\% of the centered nebulae could be bow shocks seen head- or tail-on. For the rest,
   the data disfavor explanations as remains of parental disks, supernova remnants of a
   previous companion, and dust production in stellar winds.  The existence of centered
   nebulae also at high Galactic latitudes strongly limits the global risk of coincidental
   alignments with condensations in the ISM. Mass loss during binary evolution seems a
   viable mechanism for the formation of at least some of these nebulae.  In total, about
   29\% of the IR nebulae (2\% of all OBA stars in the BSC) may find their explanation
   in the context of binary evolution.
 }
   \keywords{-- stars: early-type; emission-line, Be; rotation -- binaries -- circumstellar matter -- ISM: dust}

   \titlerunning{IR nebulae around bright massive OBA stars}
   \authorrunning{Bodensteiner et al.}

   \maketitle
%

\section{Introduction}\label{Sec:intro}
Recent observations indicate that about 70\% of all O stars are members of interacting binary systems and undergo mass transfer phases at some point of their evolution \citep{Sana2012}. Close binary interactions thus strongly influence the evolution of a large number of massive stars \citep{Podsiadlowski1992, Wellstein1999, deMink2014}. Different interaction channels are possible, consisting of tidal interactions \citep{Zahn1975, deMink2009}, mass and angular momentum transfer via Roche lobe overflow \citep[RLOF,][]{Packet1981, Shu1981}, common-envelope (CE) phase \citep{Paczynski1976, Iben1993}, and stellar merger \citep{Podsiadlowski1992, Wellstein2001, Tylenda2011}.

The outcome of such binary interactions depends strongly on the initial mass ratio and the orbital separation of the system \citep{deMink2013}. In wider systems, the mass transfer typically occurs when the donor expands after leaving the main sequence \citep[MS,][]{deMink2013}, commonly referred to as "case B mass transfer" \citep{Kippenhahn1967}. When mass is transferred during a sufficiently stable RLOF phase, the companion accretes the material and angular momentum and is spun up efficiently. In conservative mass transfer, the companion is able to accrete all material transferred from the primary, while in the non-conservative case, material is lost to the surroundings because the companion cannot accrete at the same rate \citep{Petrovic2005, Tout2012}.

After the main mass transfer event in a binary, which could be a RLOF or a \textbf{CE} evolution, the mass donor will evolve through a phase as helium star before reaching its end stage. This is best confirmed in the low-mass regime by the so-called subdwarfs. Subdwars are (mostly) helium-core burning stars that have almost completely lost their hydrogen envelope and have a narrow mass distribution centered on 0.47\,$M_{\odot}$. Subdwarfs are mainly divided into two groups, sdOs and sdBs, depending on their temperature, which can reach more than 50\,000\,K   \citep{Heber2009, Heber2016}.

In the intermediate- and high-mass regime, helium stars are expected to reach temperatures of up to 100\,000\,K \citep{Yoon2010, Langer2012, Gotberg2017}, which renders their direct detection difficult, such that only a few counterparts are known \citep{Schootemeijer2018}. However, strong evidence for their existence comes from the Be-X-ray binaries, which require an intermediate-mass helium star as neutron star progenitor \citep{VandenHeuvel1987}. For the most massive progenitors, the helium stars obtain strong winds and are easily identified as Wolf-Rayet stars with strong emission lines, such that the helium-star stage of binary evolution is again observationally well confirmed \citep{Petrovic2005}.

Depending on the initial masses of the two components, the mass donor will end up as a white dwarf (WD), 
a neutron star (NS), or a black hole (BH) \citep{Portegies1995, Han2002}. WDs are difficult to detect in massive binary systems because of their low luminosity and mass. They may, however, reveal themselves if they are interacting with their companion through mass transfer.  Such systems appear as so-called cataclysmic variables \citep{Robinson1976}, Supersoft X-Ray Sources \citep{Kahabka1997} or symbiotic binaries \citep{Kenyon1984}, in which matter is accreted 
from the companion onto the WD via an accretion disk.

If the initial mass of the primary is sufficiently high, the interaction may result in the formation of an NS \citep{Rappaport1982} or even a BH \citep{Zhang2003, Casares2014}. Because material is also accreted onto the compact object, these systems are strong X-ray sources, known as X-ray binaries (XRBs). They can be detected out to large distances because of this strong but transient X-ray emission \citep{Haberl2000, Reig2011}. When forming an NS, the primary is disrupted in a supernova (SN) explosion. The sudden mass loss caused by the ejection of the exploding star's envelope drastically changes the binary parameters and can lead to the breakup of the system. For a circular orbit, and if the NS birth kick is neglected, the system remains bound if the ejected mass is lower than half the total system mass \citep{Blaauw1961}. Owing to asymmetries in the supernova, a kick can be induced in a random direction that might hinder or support the breakup of the system \citep{Martin2009}. 

Massive stars in close binary systems with similar initial masses and short periods may even merge, forming a fast-rotating single star \citep{Podsiadlowski1992}. In addition to the fast rotation, the product of a stellar merger may show higher abundances of processed material (e.g., nitrogen), and appears rejuvenated compared to other stars in the same stellar population \citep{Glebbeek2013}. During the merging process, the huge excess of angular momentum can lead to the ejection of mass (between $2-8$\% of the total mass of the system), depending on the initial mass ratio and the evolutionary status of the stars \citep{Glebbeek2013}. In mergers of massive stars, a few solar masses might be lost, which would lead to the formation of a massive circumstellar nebula \citep{Schneider2016}. The material in such nebulae dissipates on timescales shorter than $10\,000$\,yr \citep{Schneider2016}, limiting their observability.

Systems with extreme mass ratios evolve into a CE phase directly after the onset of mass transfer \citep{deMink2013}, in which both stars orbit each other inside a single, shared envelope \citep{Ivanova2013}. CE phases are thought to be too short to lead to a substantial spin-up of the companion \citep{deMink2013, Shao2014}. However, both RLOF \citep{Packet1981} and stellar mergers \citep{Podsiadlowski1992} can result in the formation of fast-rotating stars as the angular momentum transfer spins up the companion star significantly \citep{Sills2005, deMink2013}.

Several authors have predicted that if the companion is a B star, such spin-up in a binary system will turn it into a Be star \citep{Packet1981, Rappaport1982, Pols1991}. Classical Be stars are defined as "rapidly rotating main-sequence (MS) B-type stars that are forming an outwardly diffusing gaseous, dust-free Keplerian disk" (\citealt{Rivinius2013}, but see also \citealt{Collins1987}). The process that transports matter onto a Keplerian orbit is still unknown but probably involves the combination of near-critical rotation and nonradial pulsation \citep{Rivinius1998, Baade2017}.  In the disk, the material is governed by viscosity \citep{Lee1991, Carciofi2009} and leads to prominent Balmer emission lines \citep{Rivinius2013}.

Binary population synthesis studies show that it cannot be excluded that most (or even all) Be stars are formed through mass exchange with or without subsequent merger \citep{Vanbeveren2017}.  Other possible pathways that lead to fast rotation are that the stars could be born as fast rotators \citep{Bodenheimer1995, Martayan2007} or spin up during MS evolution when angular momentum from the core is transferred onto the envelope \citep{Ekstrom2008} and, perhaps, by nonradial $g$-mode pulsations. Possibly, not all Be stars are spun up by the same process, but some or all of these models play an important role (see, e.g., \citealt{McSwain2005}, \citealt{Huang2010}). As the models are not mutually exclusive, however, a single-model-specific tracer is required to distinguish between them.

Here, we report on a search for hints of close-binary interactions by investigating the stellar surroundings at mid-IR wavelengths in different input samples of early-type stars that we visually inspect in WISE 22\,$\mu$m data. On the one hand, we investigate the surroundings of three lists of peculiar subtypes of stars, that is, Be stars, BeXRBs, and Be+sdO systems. On the other hand, to place these searches in a broader context, we also inspect the surroundings of all $\sim$\,3800 OBA stars in the Bright Star Catalogue \citep[BSC,][]{Hoffleit1991} consisting of all stars with V\,$\leq 6.5$. We characterize the stars by a literature search, explore the origin and nature of the nebulae, and discuss their formation in the context of binary evolution.


\section{Sample selection and morphological classification}\label{Sec:sample}
\subsection{Sample selection and visual inspection}\label{Sec:selection}
We investigate the surroundings of stars belonging to four input samples in WISE 22\,$\mu$m data \citep{Wright2010}. 
The main sample is taken from the BSC, a catalog of all stars with V magnitude 6.5 or brighter, and consists of all 3768 OBA stars (i.e., 1962 A stars, 1756 B stars, and 50 O stars). The BSC appears as a suitable source for this type of study. On the one hand, it contains enough stars to draw statistically significant conclusions, and on the other hand, the stars are bright enough to have been well studied in the literature. The sample is magnitude limited, not volume limited, meaning that the volume probed by O stars is much larger than that of B or A stars. More specifically, for a limiting V magnitude of 6.5, the heliocentric distance probed with an O5 star is 1500\,pc, while the distance probed by an A5 star is smaller than 100\,pc. The BSC sample serves as a comparison baseline for our population-restricted samples of Be stars, BeXRBs, and Be+sdO systems.

The Be sample consists of all known Be stars in the BSC as identified by \citet{Zorec1997} or classified as such in SIMBAD \citep{Wenger2007}, resulting in 234 candidates. Additionally, we inspected a list of 72 BeXRBs \citep{Raguzova2005}, and the Be+sdO sample of \citet{Wang2018} comprised of 5 known Be+sdO binary systems, 8 candidate, and 4 potential candidate systems. An overview of the different samples is given in Table\,\ref{Tab:samples}. 

\begin{table} \centering
\caption{Overview of the inspected samples,  where $N_\mathrm{stars}$ indicates the total number of stars in the sample (candidates are given in brackets). Fourteen of the Be+sdOs and four of the BeXRBs are also contained in the BSC sample, and all Be stars are part of it by definition. }
\begin{tabular}{lrrl}\hline \hline
Sample & $N_\mathrm{stars}$ && Source \\ \hline
Be stars & 234 && BSC / SIMBAD \citep{Zorec1997}\\
BeXRBs & 72& & \citet{Raguzova2005}\\
Be+sdOs & 5 (12) && \citet{Wang2018}\\
OBA stars & 3768& & BSC \citep{Hoffleit1991}\\ \hline
\end{tabular}
\label{Tab:samples}     
\end{table}

   \begin{figure*} \centering 
   \includegraphics[width=0.99\hsize]{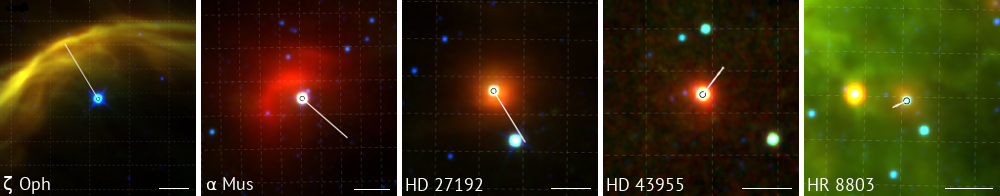}
      \caption{Representative examples for each morphological group. From left to right: aligned bow shocks, unaligned bow shocks, and centered, unresolved, and not classified nebulae. All images are RGB images where red, green, and blue correspond to the WISE 22,12, and 4.6 $\mu$m filters. The straight white lines show the proper motion corrected to the local standard of rest. Their length indicates the displacement of each star within 10,000 yr. The white bar at the bottom of each panel corresponds to 2'.}
         \label{Fig:nebuli}
   \end{figure*}

In our analysis of the BSC sample, all selected stars were inspected twice. In the first iteration, we used the WISE 22 $\mu$m images as the nebulae tend to be most prominent there. Only the brightest nebulae at 22\,$\mu$m are also visible at other wavelengths, especially at 12\,$\mu$m (WISE W3) or H$\alpha$. We checked for complementary H$\alpha$ images from the SuperCOSMOS H-alpha Survey \citep[SHS][]{Parker2005}, but because of their rarity, we focused our investigation on the WISE data at 22 $\mu$m. The stars were examined by eye in image windows of 600"\,$\times$\,600", while the flux cuts were changed manually depending on the particular field. In this step, 318 stars were identified that show a clearly extended emission in WISE W4. 

In a second and independent iteration, all selected stars were examined again. The image windows were increased to 1200"\,$\times$\,1200" to better evaluate the local surroundings. Here, we investigated
not only the 22\,$\mu$m images, but also RGB-images from the WISE data where the 22\,$\mu$m band is shown in red, the 12\,$\mu$m band in green, and the 3.4\,$\mu$m band in blue.

During this second step, about 50 stars were rejected because the suspected extended IR emission could be identified as a second star very close to the target star becoming clear in the shorter wavelength bands. For the remaining 255 cases, we estimated fluxes and sizes of the nebulae from aperture photometry. As the nebulae exhibit different morphologies, the shape of each individual nebula was used instead of a circular aperture. This shape was determined by comparing the flux values in each pixel with the local background and its standard deviation $\sigma$. A pixel is defined to belong to the nebula when the flux was 10$\sigma$ higher than the averaged background. These individual apertures were used to calculate the 22\,$\mu$m flux of each nebula.

\subsection{Morphological classification}\label{Sec:classification}
We classified the extended emission with respect to the nebular morphology and the direction of proper motion of the stars. This is especially important for bow shocks (see Sect. \ref{Sec:DiscussBowShocks}). For these latter nebulae, we defined a misalignment angle $\alpha$ as the unsigned angle between the direction of proper motion of the star and the bow shock vector, connecting the position of the star with the apex of the bow shock.

We defined the following morphology groups (see Fig. \ref{Fig:nebuli}), where the number in parentheses gives the final number of nebulae classified in this group:
\begin{itemize}
\item \textbf{Aligned bow shocks (34):} nebulae that have a sickle-like shape, a clear offset between the star and the nebula, and a misalignment angle $\alpha \leq 20^{\circ}$ (within the errors of the proper motion vector).
\item \textbf{Unaligned bow shocks (60):} nebulae that have the sickle-like shape typical of bow shocks, a clear offset between the star and the nebula, but a misalignment angle $\alpha \geq 20^{\circ}$. This means that the apparent motion of the star is not aligned with the opening angle of the nebula.
\item \textbf{Centered (52):} clearly extended and not sickle-shaped nebulae, where the star is situated within the extent of the nebula. The shapes are mostly roundish, elliptical, diffuse, or asymmetric, with a wide range of irregular morphologies. 
\item \textbf{Unresolved (72):} nebulae for which the image resolution in WISE is not good enough to classify the shape, that is, sickle-like shapes cannot be distinguished from round or circular nebulae. The lack of resolution means that no offset between nebula and star can be discerned. 
\item \textbf{Not classified (37):} nebulae that cannot be unambiguously classified into one of the other groups because of field crowding, background contamination, or both. 
\end{itemize}

After defining the morphology groups, we checked for  24\,$\mu$m data from the Multiband Imaging Photometer for Spitzer \citep[MIPS,][]{Rieke2004} for all 255 nebulae. MIPS has a point spread function (PSF) size of 6'', thus twice the resolution as WISE \citep{Rieke2004}. Unfortunately, Spitzer did not perform an all-sky survey, so sky coverage is sparse and only 61 of the 255 nebulae found in WISE were observed by MIPS as well. In 25 of the 61 cases, the higher-resolution MIPS data allowed us to reclassify the nebular shape and estimate a misclassification error for the nebulae without MIPS data coverage. 

Nineteen nebulae initially classified as aligned or unaligned bow shocks were observed with MIPS, but only in one case did the higher resolution lead to a reclassification. This indicates that the bow shock classification is robust at lower resolution as well, with a misclassification error of roughly 5\%. For 23 nebulae classified as centered, higher resolution MIPS data were available, which in five cases revealed a previously hidden bow shock. This implies a misclassification rate for centered nebulae of 22\%. Thirteen nebulae classified as unresolved in WISE have additional MIPS data. Six of the cases showed an offset in MIPS and revealed a bow shock not seen in the inspection of the WISE data, while three cases could be reclassified as centered nebulae, adding up to a reclassification in 70\% of the cases. Five of the six not classified nebulae (i.e., 83\%) observed by MIPS were reclassified into other groups. As expected, the classification of unresolved and not classified nebulae could be strongly improved by higher resolution data.

\begin{figure}\centering
   \includegraphics[width=0.99\hsize]{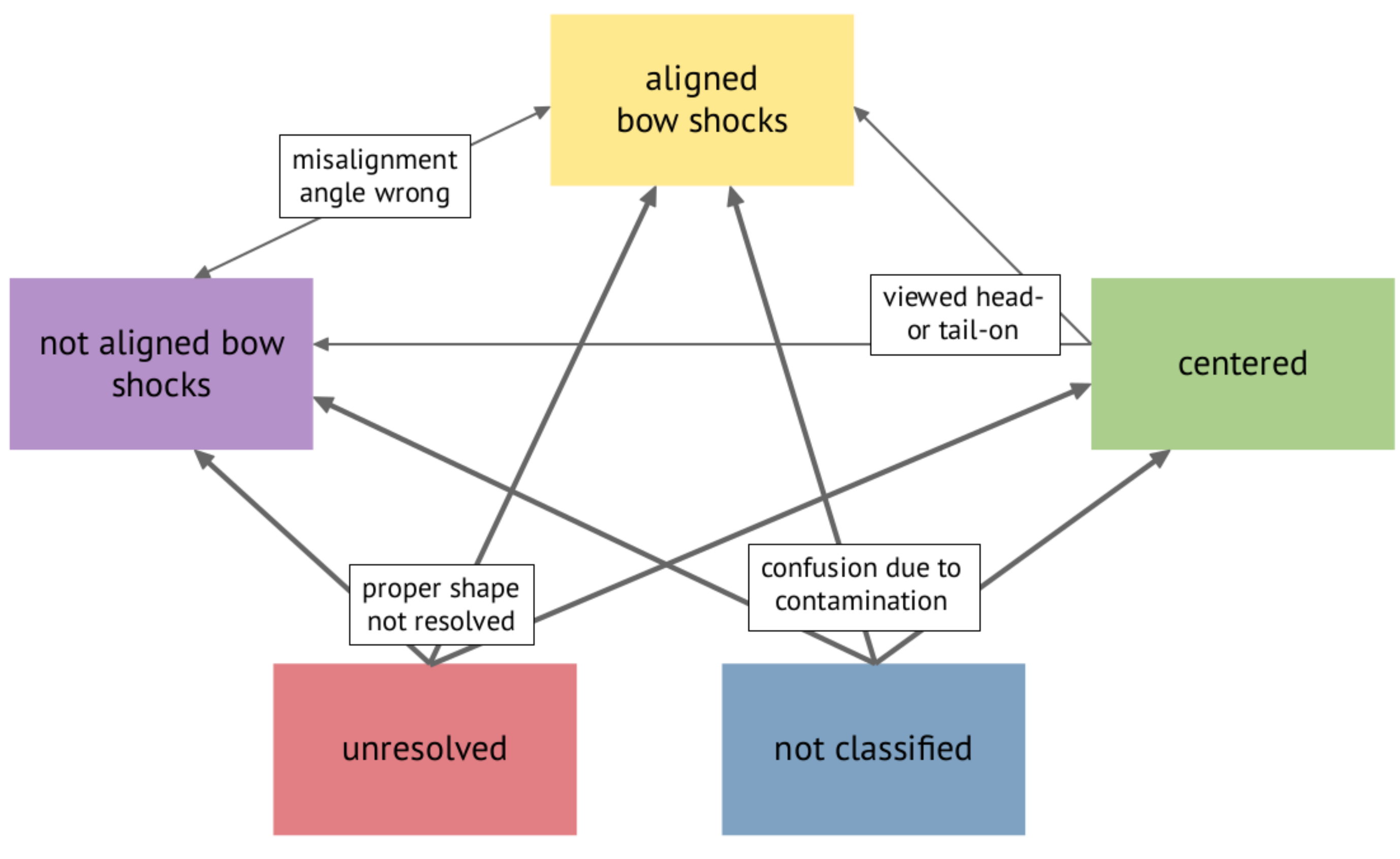}
      \caption{Sketch of the possible misclassification paths between the defined morphology groups.}
         \label{Fig:miss}
   \end{figure}

Figure \ref{Fig:miss} illustrates the possible misclassification paths arising from insufficient resolution. Additionally, bow shocks might have been classified as unaligned bow shocks and vice versa if the proper motion values taken from the literature were incorrect. This might be the case because for most of the stars, only Hipparcos \citep{VanLeeuwen2007} proper motions are available. This error can be reduced with the 2nd Gaia data release, still pending at the time of
writing, which should feature better proper motion values for the BSC stars. 

For the statistical evaluation of the sample we set up a database containing literature information as well as the parameter values measured by us for all stars we found to be associated with nebulae. It is given in Tables \ref{Tab:lit_table} and \ref{Tab:meas_table} and can be accessed via the Centre de Données astronomiques de Strasbourg (CDS).

Table \ref{Tab:lit_table} gives an overview over the parameters we gathered from the literature, including identifiers, coordinates, HD numbers, V-band magnitudes, and spectral types, taken from SIMBAD. 
Additionally, we provide parallaxes $p$, proper motions $\mu,$ and radial velocities $v_{\mathrm{rad}}$ with their respective errors for all available stars. 

Table \ref{Tab:meas_table} lists the parameters we derived. These include the final morphological classification we assigned to each nebula after inspection of the MIPS data and information about whether the star is a known Be star, has a nebulous spectrum, or is a known binary. We calculated distances $d$ and measured angular sizes and total fluxes of the nebulae. For bow shocks and unaligned bow shocks, we give angular and linear distances $R(0)$ between the position of the star and the apex of the bow shock and misalignment angles $\alpha$ as defined above. Additionally, we calculated space velocities of the stars from literature values for distance $d$, proper motion $\mu,$ and radial velocity $v_{\mathrm{rad}}$ via 

\begin{equation}\label{Eq:vspace}
v_{\mathrm{space}} = \sqrt{v_{\mathrm{rad}}^2 + \left(4.744 \cdot \mu \cdot d \right)^2} \, .
\end{equation}
Because of large errors both in distance and proper motion, we only give space velocities for stars for which the error in each individual measurement is < 30\%.

\section{Results} 
\subsection{BSC stars with nebulae}\label{Sec:BSCresults}
We find that 255 of all OBA stars in the BSC (i.e., $\sim$7\%) have associated extended IR nebulae, visible in WISE data at 22\,$\mu$m. Of these, 14 are associated with O stars, 234 with B stars, and 7 with A stars. Because of the strong dependency on spectral type of the stellar content of the BSC, the fraction of nebulae, however, declines strongly toward later spectral types: 28$\%$ of all O stars, but only 13$\%$ of the B and 0.4$\%$ of the A stars are associated with IR nebulae. Of the 7 A stars with nebulae, 3 are supergiants with luminosity classes Ia and Ib (HD\,21389, HD\,20041, and $\iota^2$\,Sco) while 1 ($\xi$\,Cep) is classified as spectroscopic binary in SIMBAD.

Absolute and relative abundances of the morphological types of IR nebulae defined in Sect. \ref{Sec:classification} are provided in Table \ref{Tab:Stats_shape}. Fig.\,\ref{Fig:nebuli} shows a representative example of each morphological category. About 40\% of all objects belong to the ambiguous groups of unresolved and not classified nebulae. The aligned bow shocks are the rarest objects in this sample, while 52 nebulae were classified as centered. Nebulae classified as centered might actually be bow shocks viewed head- or tail-on that appear roundish and centered on the star as a result of projection effects (see Sect. \ref{Sec:DiscussCentered}).

\begin{table*}\centering
\caption{Classification scheme based on the shape of the nebulae associated with BSC stars, the misalignment angle $\alpha$, and the spatial offset between the star and the nebula. Columns 5 and 6 give the absolute number $N_\mathrm{neb}$ and the relative frequency $f_\mathrm{neb}$ of nebulae belonging to the respective groups. To estimate counting errors, we use the binomial error formula $\Delta f = \sqrt{(f(1-f) \ N}$ (not taking into account the misclassification errors). Here, we give the updated numbers and fractions after the reclassification based on the MIPS data. The last column indicates the potential formation mechanism (see Sect. \ref{Sec:DiscussBSC}).} 
\begin{tabular}{llclccl} \hline \hline
Classification & Shape & $\alpha$ & offset &$N_\mathrm{neb}$ & $f_\mathrm{neb}$ & Potential formation mechanism \\ \hline 
aligned bow shock &sickle-like & $\leq 20^{\circ}$ & yes & 34 & 13 $\pm$ 1 \%  & high-velocity star \\
unaligned bow shock & sickle-like &$\geq 20^{\circ}$& yes & 60 & 24 $\pm$ 1 \% & motion of ISM \\
centered & round/elliptical/diffuse & - & no & 52 & 20 $\pm$ 1 \% & RLOF in binary system \\
unresolved & not resolved &- & no & 72 & 28 $\pm$ 1 \% & - \\
not classified & ambiguous & - & -&37 & 15 $\pm$ 1 \% & - \\ \hline
\end{tabular}
\label{Tab:Stats_shape}
\end{table*}

\begin{figure}\centering
   \includegraphics[width=0.99\hsize]{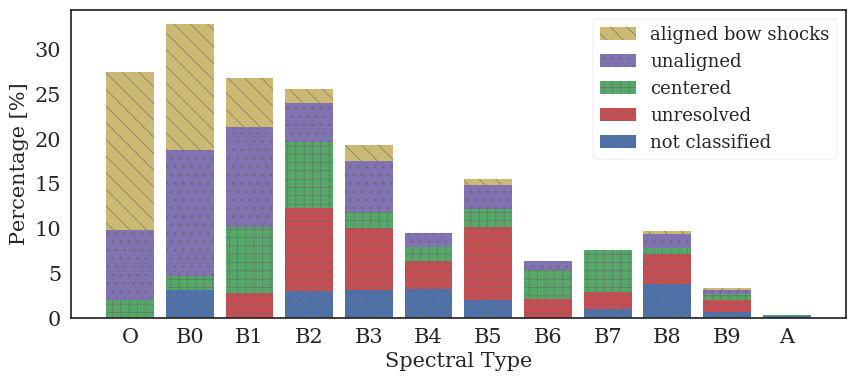}
      \caption{Associations of different morphology groups with spectral types. Because of their low numbers, O and A stars are not divided into subtypes. The fractions are given with respect to the total number of stars of the respective spectral type in the BSC.}
         \label{Fig:spec_histo}
   \end{figure}

In Fig. \ref{Fig:spec_histo} we show the association frequency of morphology types with spectral subtypes. The morphology types exhibit different dependencies on spectral type: The aligned bow shocks are the main reason for the overall decline in spectral type: while the fraction is highest for O stars, it decreases strongly toward later spectral types. Unaligned bow shocks are also less frequent around later-type stars, but the decline is not as strong as for the aligned bow shocks. 
Centered nebulae can be found at all spectral types with a peak around B1-B2. 

\begin{figure}\centering
   \includegraphics[width=0.99\hsize]{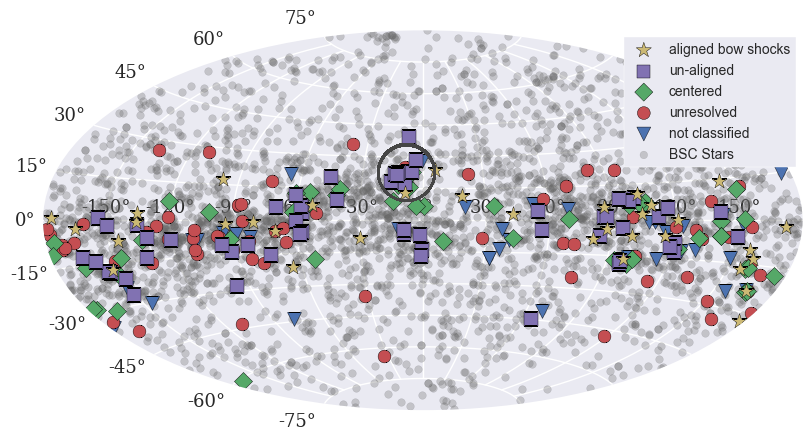}
      \caption{Distribution of all O, B, and A stars in the BSC over the sky (gray dots, transparent for better visibility). The BSC stars associated with different morphological types of IR nebulae are identified with the symbols explained in the legend. The Ophiuchus star-forming region is marked by a circle.}
         \label{Fig:Aitoff}
   \end{figure}

As demonstrated in Fig. \ref{Fig:Aitoff}, we find an accumulation of nebulae along the Galactic plane as well as tracing Gould's Belt. Gould's Belt is a ring-like region tilted with respect to the Galactic plane by 16 - 20$^{\circ}$, in which large OB associations are located, that is, the Orion, Ophiuchus, and Cygnus star-forming regions \citep{Poppel1997}, which contain many young O and B stars and atomic and molecular gas \citep{Perrot2003}. However, the association with IR nebulae is not restricted to a narrow strip around the Galactic plane, and nebulae are also
found at higher latitudes.

\begin{figure}\centering
   \includegraphics[width=0.99\hsize]{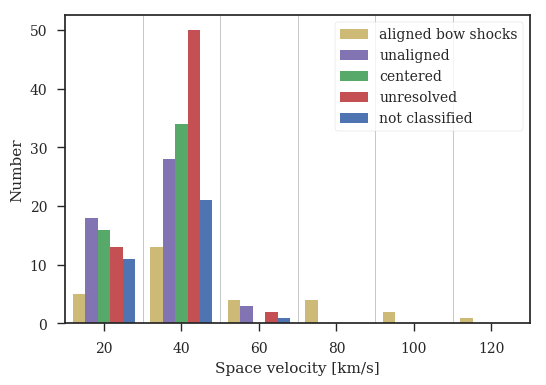}
      \caption{Distribution of space velocities of the BSC stars calculated from Eq.\,\ref{Eq:vspace}. The colors correspond to the different morphological groups as indicated in the legend.}
         \label{Fig:vspace}
   \end{figure}

Figure \ref{Fig:vspace} shows the space velocity distribution for each of the morphological categories. As mentioned in Sect. \ref{Sec:classification}, we only calculated space velocities for the stars with individual errors in $d$, $\mu$, and $v_{\mathrm{rad}}$ <\,30\%. The total space velocity of stars associated with aligned bow shocks is on average higher than the velocity of the stars associated with other morphological types: all nebulae associated with stars with velocities higher than 60\,km/s are aligned bow shocks. On the other hand, there are also a few aligned bow shocks around low-velocity stars. Centered nebulae can be found around stars with velocities of up to 40\,km/s, while unresolved and not classified nebulae are associated with stars with velocities up to 60\,km/s.

\subsection{Subsamples}
Here we investigate the stars included in the three subsamples defined in Sect. \ref{Sec:selection} for associations with IR nebulae. We employed the same inspection approach as for the BSC sample described in Sect. \ref{Sec:selection}. 
\subsection*{Be stars}
Of the 234 inspected Be stars, 34 (i.e., 15$\%$) are associated with IR nebulae. This can be compared to the fraction of normal B stars (excluding the Be stars) with associated nebulae, which is 13$\%$.
Among these 34 nebulae, there are 4 aligned bow shocks, 7 unaligned bow shocks, and 11 centered nebulae. This means that in total, 1\% of all the Be stars are associated with aligned bow shocks, 3\% with unaligned bow shocks, and 5\% with centered nebulae.
 
More B stars in the BSC might be Be stars as the Be phenomenon is transient and might not be detected as such in observations with limited time coverage. This would influence the ratio between B and Be stars in the BSC and thus the comparison between the fraction of B and Be stars associated with nebulae. Eight Be stars with nebulae were also observed with MIPS. However, in none of these cases did the additional higher resolution data change the morphological classification.
\subsection*{BeXRBs}
We investigated a total of 72 BeXRBs, among which we found 4
associations with IR nebulae and 15 uncertain cases. As BeXRBs are typically detected through their strong X-ray emission (see Sect.\,\ref{Sec:intro}), they are observed to much larger distances than the typical B stars in the BSC sample. Only 4 BeXRBs are within the optical brightness limits of the BSC. Two of these are associated with an IR nebula: $\gamma$\,Cas (not classified) and $\mu^2$\,Cru (centered). The detection of the other 2 nebulae around BeXRBs, where the stars are too faint to be included in the BSC, should be considered with caution, as the stars are probably at much larger distances. Hence the risk of chance alignments with ISM structures increases,  especially for objects relatively close to the Galactic plane. This is the case for $\gamma$\,Cas and $\mu^2$\,Cru: they are within $\pm 6^{\circ}$ of the Galactic plane. 
Only $\mu^2$\,Cru was observed by MIPS, but this did not change its morphological classification.
\subsection*{Be+sdOs}
Of the 5 known Be+sdOs and the 12 candidate and potential candidate systems, 3 (HD\,41335 and HD\,214168, both centered, and $\phi$ Per, not classified) are associated with IR nebulae and 1 case is uncertain (59 Cyg, not classified). The clear detections are also part of our BSC sample. Only for HD\,41335 are MIPS data available that confirm the classification as centered nebula.

The findings for these three subsamples are summarized in Table \ref{Tab:Stats_sub}. Because of incompleteness of the input catalogs, the absolute numbers can only be lower limits, especially for the Be+sdO systems, which are difficult to identify. Absolute and relative numbers are of course subject to classificational uncertainties, as described in Sect. \ref{Sec:classification}.  

\begin{table*}[h!]\centering
\caption{Number of nebulae classified in the morphology groups defined in Sect. \ref{Sec:classification} and associated with stars in the subsamples. The first column refers to the classification in shape. The subsequent columns give absolute number and relative frequency of nebulae belonging to the Be star, BeXRB, and Be+sdO samples, respectively. Counting errors are estimated with the binomial error formula $\Delta f = \sqrt{(f(1-f) \ N}$. The last column indicates the potential formation mechanism (see Sect. \ref{Sec:DiscussBSC}).} 
\begin{tabular}{lrcccccccccl} \hline \hline
Classification & & \multicolumn{3}{c}{Be stars} & \multicolumn{3}{c}{BeXRBs} & \multicolumn{2}{c}{Be+sdOs} \\ \hline
& & $N_\mathrm{neb}$ & $f_\mathrm{neb}$ && $N_\mathrm{neb}$ & $f_\mathrm{neb}$& &$N_\mathrm{neb}$ & $f_\mathrm{neb}$ & & Potential formation mechanism \\ \hline
aligned bow shocks && 4 & 12 $\pm$ 2\% & & - & - & & - & - & & high-velocity star\\
unaligned bow shock & & 7 & 21 $\pm$ 3\% & & - & -  & & - & - & & motion of ISM \\
centered & & 11 & 32 $\pm$ 3\% & & 1 & 25 $\pm$ 5 \% & & 2 & 67 $\pm$ 11\% & & RLOF in binary system\\
unresolved & & 6 & 18 $\pm$ 2\% & & - & - & & - & - & & -\\ 
not classified & & 6 & 18 $\pm$ 2\% & & 3 & 75 $\pm$ 5 \% & & 1&  33 $\pm$ 11\% & & - \\ \hline
\end{tabular}
\label{Tab:Stats_sub}
\end{table*}

\section{Discussion}
\subsection{BSC sample}\label{Sec:DiscussBSC}
\subsubsection{Unresolved and not classified nebulae}
As described in Sect. \ref{Sec:classification}, about 70\% of the nebulae unresolved by WISE may be resolved with a twice narrower PSF. Accordingly, about 47 or more of the total of 68 unresolved nebulae without MIPS data might be reclassified at higher resolution. Applying corresponding statistical corrections, we estimate that about 31 might be bow shocks, 16 might be centered nebulae, and 21 would still be unresolved. Their individual genuine character can, however, only be derived from higher-resolution data. The large number of unresolved nebulae with a space velocity > 30\,km/s additionally indicates that a majority of them could be hidden bow shocks. In addition to the unresolved nebulae, we do not take into account the not classified nebulae in the following as in the context of this work no information can be drawn from them. 

\subsubsection{Aligned bow shocks}\label{Sec:DiscussBowShocks}
Bow shocks develop when the relative velocity between a star with a strong stellar wind and the local ISM is supersonic \citep[see, e.g.,][]{VanBuren1988, Comeron1998, DelValle2012}. The size and brightness of a bow shock are defined mainly by three parameters: Theoretical modeling shows that a higher relative velocity as well as a higher density of the ambient ISM leads to smaller bow shocks, while their size increases with increasing wind strength and the thereby induced mass loss \citep{Meyer2014, Meyer2016}. Stellar winds are generally stronger in more massive stars, leading to the expectation that bow shocks are more common and prominent around earlier-type stars. In the case of later-type (i.e., late-B and A-type) stars the radiation pressure on dust grains might lead to the formation of a bow shock \citep{VanBuren1988, Ochsendorf2014}.

Historically, the formation of a bow shock was attributed to a high stellar velocity with respect to its surroundings \citep{VanBuren1995, Kaper1997}. The stars producing such bow shocks are called runaway stars and typically move at a high peculiar velocity  $>30\mathrm{km\,s}^{-1}$ \citep{Blaauw1961, Gies1986}. The runaway population is dominated by O stars \citep{Stone1991} that are observed outside of star-forming regions, indicating that they moved away from their place of formation \citep{Silva2011}. 
Two possible mechanisms may lead to a high peculiar velocity: (i) expulsion from an SN-disrupted binary \citep[see Sect. \ref{Sec:intro} and][]{Blaauw1961, Tauris1998}, and (ii) encounters between a close binary system and a tertiary star or another binary system in the parental cluster, which can lead to the ejection of one of the stars \citep{Poveda1967}. Observations indicate that both mechanisms are playing an important role \citep{Hoogerwerf2000}. 

The alignment in our first morphological group between the motion of the stars and the sickle shape of the associated bow shocks implies that the associated stars are high-velocity stars. The steep decline with spectral type (see Fig. \ref{Fig:spec_histo}) agrees with the theoretical prediction that bow shocks are stronger around earlier-type stars as they have much stronger winds. A majority of the A stars with bow shocks are either supergiants or in a binary system. While MS A stars in general have no winds of any relevance, supergiants probably do have a wind strong enough for the formation of a bow shock \citep{McCarthy1997}. The recent observation of a bow shock around the red supergiant star IRC-10414 \citep{Gvaramadze2014} further supports this. 

Additionally, the decline with spectral type reflects the fact that the population of high-velocity stars is dominated by early-type stars. Most aligned bow shocks in our sample are associated with stars that have high space velocities (i.e., up to 120\,km/s, see Fig. \ref{Fig:vspace}). A small number of outliers has $v_{\mathrm{space}} < 30$\,km/s, probably indicating that in these cases the ISM motion dominates the relative motion and is by chance aligned with the stellar motion (see next subsection).

Because the velocity gain of at least some runaway stars arises from their binary origin, the 34 stars with bow shock nebulae (i.e., 13\% of all the stars with nebulae) can be placed in the context of binary evolution.
Additionally, by using aligned bow shocks as indicators of high-velocity stars, we can crudely estimate their fraction in our limited sample of 50 O stars. We find 9 associations between O stars and aligned bow shocks, that is, almost 20\%. This agrees well with other studies that found a high-velocity star fraction between 10\% \citep{Blaauw1961, Gies1986} and 27\% \citep{Tetzlaff2010}.

\subsubsection{Unaligned bow shocks}\label{Sec:DiscussBSNA}
Recent observational studies showed that bow shocks are not restricted to stars that move with high space velocities \citep{Povich2008, Kobulnicky2016}. Large-scale motions within the ISM can also be responsible for high relative velocities between star and ISM \citep{Povich2008}. Our morphological group of unaligned bow shocks supports this finding: stellar motion with respect to the ISM cannot be the driving force in the formation of these bow shocks. 

As described in Sect. \ref{Sec:BSCresults}, unaligned bow shocks are associated with all spectral types we studied. This agrees well with theory: The main part in this interaction is the motion of the ISM, not stellar motion. As described before, O stars are favored as nuclei of bow shocks for two reasons: their strong winds, and high velocities. While the role of the wind should not depend much on the stellar space velocity with respect to the ambient ISM, the velocity threshold for the occurrence of bow shocks is obviously lower if the ISM also contributes much to the relative velocity between star and ISM.  It is therefore not surprising that the fraction of early-type B stars with unaligned IR nebulae, which have weaker winds, is larger than that of aligned bow shocks (Fig. \ref{Fig:spec_histo}). In B stars in general, the formation of a bow shock depends more strongly on the ISM contribution to the relative velocity. This is confirmed by the distribution of space velocities of the stars with unaligned bow shocks: as shown in Fig. \ref{Fig:vspace}, the space velocity of these stars is in general significantly lower than for the aligned bow shocks, that is, mostly < 40\,km/s.

Unaligned bow shocks can be used as tracers of large-scale ISM motions \citep{Kiminki2017}. Such motions are common in star-forming regions where they are created by the copious winds of hot stars \citep{Kudritzki2000} and multiple previous SN explosions \citep{DeAvillez2010}. The Ophiuchus star-forming region, for example, clearly shows an overabundance of unaligned bow shocks (encircled in Fig. \ref{Fig:Aitoff}). Accordingly, the defining link between bow shocks and high stellar velocity is relaxed. The basic physics of aligned and unaligned bow shocks, however, remains the same because of the roughly spherical symmetry of the stellar winds.  

\subsubsection{Comparison to Kobulnicky et al. 2016}
\citet{Kobulnicky2016} carried out a comprehensive visual inspection of existing space-based imaging IR surveys, and presented a new catalog of candidate bow shock nebulae containing 708 objects. In a follow-up paper, they performed aperture photometry and derived spectral energy distributions (SEDs) for all the objects \citep{Kobulnicky2017}. For candidate bow shocks, the authors imposed the following classification criteria: the nebulae must have an arc shape, a high degree of symmetry, and at least one prominent star located near the axis of symmetry.  Subsequently, they classified the detected nebulae into four groups depending on their environment: bow shocks located in apparently isolated environments (70\% of all nebulae), facing H\,II regions, facing bright-rimmed clouds, and located within H\,II regions. They called the latter "in situ" bow shocks, indicating that these bow shocks do not originate from the fast motion of a high-velocity star, but from ISM motion. 

Reliable proper motion data are only available for $\sim 20\,\%$ of their stars. In one-quarter of the latter, the position angle of the bow shock is found to be aligned with the direction of the proper motion. This value is lower than the fraction of aligned bow shocks (34) of all bow shocks (94) in the BSC, which is 36\%. It has to be noted, however, that Kobulnicky et al. defined bow shocks as aligned if the misalignment angle $\alpha$ is $\leq 45^{\circ}$, while our cutoff is $\leq 20^{\circ}$. Their derived fraction of aligned bow shocks is nevertheless lower than ours,
which might at least partly be explained by the fact that, in contrast to our study, their proper motion vectors were not corrected for the local standard of rest.

The approaches used by Kobulnicky et al. and by us are complementary to each other:  Kobulnicky et al. searched for IR nebulae with bow shock morphologies, classified them according to their environment, and then studied their association with stars, whereas we searched all OBA stars in the BSC for an associated IR nebula. Therefore, our analysis permits inferences to be made not only about the incidence of strong interactions between stellar winds and the ISM, but also about ISM-independent processes that may lead to the formation of IR nebulae associated with OBA stars. Moreover, Kobulnicky et al. inspected archival MIPS images mostly covering the inner Galactic plane at latitudes between $|b| \leq 1^{\circ}$, and complementary WISE data at latitudes up to $|b| \leq 2^{\circ}$. In contrast, our study covers the whole sky (therefore it mostly depends on the lower resolution WISE data), but is not restricted to bow shocks. We start from a (magnitude-limited) sample of bright stars of spectral types O, B and A and search for associated nebulae of any morphology.

A comparison between the nebulae found by Kobulnicky et al. and our sample yields only a small overlap of 11 stars (5 aligned and 6 unaligned bow shocks; by definition of the Kobulnicky et al. sample, there are no common objects with other structures). Of these 11, 10 fall into Kobulnicky et al.'s category of "isolated" environments (as the majority of their sample, i.e., 70\,\%), and 1 unaligned bow shock is "facing an H II region". As mentioned in Sect. \ref{Sec:BSCresults}, we only have 14 O stars associated with IR nebulae in our sample. It is thus interesting to note that 5 of the 11 stars in the overlap are associated with O type stars and the rest are associated to B0 stars. A possible reason for the restriction of common objects to very early-type stars could be the clustering of these objects around the Galactic plane. The overall small overlap could be due to the different input data: while our search includes all OBA stars within a certain brightness limit in the whole sky, Kobulnicky et al. focused on a small region close to the Galactic plane without taking into account the spectral type or brightness of the associated star. 

\subsubsection{Centered nebulae}\label{Sec:DiscussCentered}
Our third morphology group consists of all nebulae in which the emission is centered on the star. The shape of these nebulae is circular, elliptical, or diffuse. Possible explanations for the formation of IR nebulae without obvious offset from the associated star include the following. 
\begin{itemize}
\item Some of the centered nebulae could be bow shocks viewed head- or tail-on. Their associated stars should exhibit small proper motions, that is, low tangential velocities, but fast radial velocities. We calculated space velocities for 50 of the 52 centered nebulae (the remaining 2 were omitted because of high uncertainties in distance and proper motion, see Sect. \ref{Sec:classification}). Of these 50, only 2 stars have a high tangential velocity $\mathrm{v_{tan}}$ > 30\,km/s, while 2 others have radial velocities $\mathrm{v_{rad}}$ > 30\,km/s. Additionally, the abundance of associations with centered nebulae depends on spectral type. A peak is visible at spectral types B1/B2, implying that not all centered nebulae are bow shocks viewed head- or tail-on. 

In Appendix A we compare the distribution of apex distances of bow shocks to the distribution of radii of centered nebulae. We calculated the radii from the physical sizes of the nebulae assuming that all nebulae have a circular shape. As the centered nebulae are not all spherically symmetric, and because the measured apex distance is a projected quantity, we do not expect the underlying distributions to be similar. We tested the similarity of the two distributions by means of a Kuiper test, which indicated that we cannot distinguish between them at a 2$\sigma$ level.

\citet{Acreman2016} computed synthetic IR emission maps for bow shocks viewed at inclination angles 30$^{\circ}$, 60$^{\circ}$ , and 90$^{\circ}$ smoothed to the resolution of WISE W4 (see their Fig.\,3). The authors found a "noticeable asymmetry" of the nebula "with a peak ahead of the star" already at an inclination of 30$^{\circ}$ \textbf{that becomes} a more clearly identifiable bow shock at higher inclinations. We thus assume that bow shocks seen under inclination angles $\leq$\,30$^{\circ}$ appear spherical with WISE. Assuming a continuous distribution of bow shock orientations in the sky, the fraction of bow shocks viewed head- or tail-on is given by the integral of a random distribution of angles from $0^{\circ}$ to $30^{\circ}$, that is, $1-\cos(30^{\circ}) = 13\%$. This implies that at least 13\% of all bow shock nebulae would be classified as centered nebulae because of their head- or tail-on orientation. This indicates that the 94 bow shock nebulae we found correspond to 87\% of the total bow shock population (i.e., 108 nebulae), and that at least 11 centered nebulae should be bow shocks viewed head- or tail-on. Applying this additional possible misclassification to all centered nebulae statistically reduces their number from 52 to at most 41, that is, 16\% of all nebulae.

\item The IR emission could arise in the remains of the parental disk in which the star formed. However, these disks are transient structures with lifetimes on the order of a few million years. The lifetimes steeply decrease with the mass of the star \citep[see, e.g.,][]{deMarchi2013, Pfalzner2014} because the strong stellar irradiation leads to disk dissipation \citep{Haisch2001}. This is much shorter than the typical MS lifetime of early-type stars \citep[e.g., $\sim 2\cdot 10^8$yr for B5V,][]{Ekstrom2012}. Moreover, protostellar disks are smaller than $1000\,\mathrm{AU}\,\approx 5\cdot 10^{-3}\,\mathrm{pc}$ \citep{Williams2011}, whereas the typical radii of the observed IR nebulae of $0.5\,\mathrm{pc}$ are 100 times larger.

\item The nebulae could be the remnants of a supernova explosion of a companion star. For a typical radius of $0.5\,\mathrm{pc}$ of the IR nebulae in question and typical expansion velocities of supernova remnants \citep[5,000\,km/s,][]{Hayato2010}, putative SN explosions would have taken place 100 years ago, i.e. during historical times. A typical absolute luminosity of core-collapse SNe is $-17$\,mag in V \citep{Richardson2014}, which is about 14 mag brighter than that of a B1 V star \citep{Wegner2006}. Therefore, even the faintest B1V star in the BSC at $m_V$\,=\,6.5 would have shone at $M_V\,=\,-7.5$ around maximum, which is 10,000 times brighter than Sirius. There is no record of such events.

\item The IR emission could arise in surrounding dust that is heated by the star. This could either be dense ISM clouds that are independent of the star, or dust formed by the star itself. Additional observations of one peculiar nebula around a classical Be star, HD\,303056 (classified as B3 in SIMBAD), indicated a nebular density of n$_{\mathrm{neb}}$=56$\pm$5\,cm$^{-3}$ (Bodensteiner, Master thesis, unpublished; Bodensteiner et al., in prep.). The typical density of the ISM, however, is n$_{\mathrm{ISM}} \approx 1\, \mathrm{cm}^{-3}$ \citep{Ferriere2001}. Therefore, if this dust is unrelated to the star and part of the ISM, the density of the latter would have to be accidentally increased by nearly two orders of magnitude, which seems very unlikely. 

Figure \ref{Fig:Aitoff} demonstrates that the association of nebulae is not restricted to a narrow strip around the Galactic plane. It extends to higher Galactic latitudes, showing that such nebulae do not only occur in regions of dense ISM where the probability of both physical and apparent associations is highest. This suggests that some of the IR nebulae without bow-shock appearance may not require the ISM, but are formed by the associated stars in some other way.

Two basic conditions must be fulfilled in stars in order to produce dust: the temperature of the material must be below the dust condensation temperature of around $1500$\,K \citep{Hoefner1998}, and the density must be high \citep{Andersen2007}. These conditions are fulfilled for example in AGB stars \citep{Ferrarotti2006}, or in evolved massive stars like luminous blue variables (LBVs) or Wolf-Rayet (WR) stars \citep{Toala2015}. 

The mass-loss rate of AGB stars is between $10^{-8}$ and $10^{-4}\,M_{\odot}\,\mathrm{yr}^{-1}$ \citep{Habing1996, Kemper2001}, leading to the formation of circumstellar nebulae \citep{Wood2004, Suh2004}. LBVs show stellar outbursts and episodes of high mass loss during which the stars lose significant fractions of their initial mass \citep{Humphreys1994, Smith2014}. The wind mass loss rates of LBVs is typically  $10^{-5}$ to  $10^{-4}\,M_{\odot}\,\mathrm{yr}^{-1}$ \citep{Lamers1988, Smith2004, Groh2009}. Several dusty nebulae are known around LBVs \citep{Weis2003, Gvaramadze2009}. 

The mass loss in MS B stars is in general very small compared to these numbers. B0 stars can have a mass loss of up to $10^{-9}\,M_{\odot}\,\mathrm{yr}^{-1}$. This number, however, decreases strongly with effective stellar temperature. Below 15\,000\,K, no homogeneous line-driven wind is possible \citep{Krticka2014}, making the mass loss completely ineffective in terms of the formation of such nebulae.

\item The IR nebulae could be formed by non-conservative mass transfer during RLOF or in a stellar merger. Depending on the type of interaction, the remaining star would be effectively single (i.e., when the two components have merged) or have a compact companion (i.e., a WD, sdO, NS, or even a BH). As the least massive of these compact companions, WDs and sdOs, are hard to detect, many of them, if they exist, may have remained undetected.
The nebulae would then become visible because the ejected material is heated by the absorption of photons from the OB star, and in the case of RLOF, additionally by the compact companion. These compact companions, especially sdOs, have very high temperatures and thus emit around 100 times more highly energetic photons than an early-type star (the number of ionizing photons for an sdO is $10^{48}$\,s$^{-1}$ \citep{Gotberg2017}, while a B1 type star with an effective temperature of $\sim$\,26\,000\,K typically emits $10^{46}$\,s$^{-1}$ \citep{DiazMiller1998}. Additionally, the temperature of sdOs remains high for a long time, increasing the visibility period of such nebulae. 
\end{itemize}

This, perhaps still incomplete, multitude of possibilities shows that centered nebulae do not have a unique straightforward explanation. We can exclude that they are the remains of the parental disk or remnants of a previous SN explosion of a possible companion star, however. Because of the negligible mass loss of the stars in comparison to known dust-producing stars like AGBs, LBVs, or WR stars, it also seems unlikely that single stars produced the dust. Finally, some of the centered nebulae could be bow shocks viewed head- or tail-on. 

In conclusion, we propose that the formation of most of the remaining centered nebulae (i.e., 16\% of all nebulae) is related to binary interactions by non-conservative mass transfer either during RLOF or a stellar merger. 

\subsection{Subsamples}

\subsubsection{Be stars}
The search for nebulae within the Be subsample yields three main results. First, we find a similar fraction of nebulae around Be stars as around normal B stars. This indicates that the presence of a circumstellar disk, which is the main characteristics of a Be star, is not significantly causal for the presence of an IR nebula. 

Second, the fraction of centered nebulae around Be stars is marginally higher than around normal B stars. While 11 of the 34 Be stars, that is, 32\%, are found to be associated with centered nebulae, this holds only for 39 of the 200 normal B stars (excluding Be stars) in the BSC, that is, 20\%. The B stars in the BSC also include supergiants with strong winds.  Therefore, if the distribution of space velocities of B and Be stars is about the same, the smaller percentage of Be stars with centered nebulae confirms that winds are not generally important for the formation of centered nebulae. 

Third, four Be stars are associated with aligned bow shocks. \citet{Berger2001} found the fraction of high-velocity stars among Be stars to be slightly higher than the fraction among normal early-type B stars (i.e., 3\%-7\% in comparison to $\approx 2\%$). They concluded that the existence of high-velocity Be stars indicates that at least some of the Be stars are formed in binaries. This is in agreement with our conclusion: the four Be stars with bow shocks probably gained their high velocity in previous binary interactions. 

In total, this indicates that 11 + 4 out of 234 inspected Be stars, that is, 6\%, are associated with nebulae that can be linked to binary interactions. In comparison, we found 39 centered nebulae + 20 aligned bow shocks associated with the 1522 normal B-type stars in the BSC, implying a fraction of 4\%. The absolute numbers of star-nebula pairs and the corresponding fractional difference between Be and B stars, 6\% vs.\ 4\%, are small, however. Therefore, on the basis of our data, we cannot claim to see strong evidence that binarity is the physical cause of the Be star phenomenon.

The first detection of an elongated, asymmetric IR nebula in WISE data at $22\,\mu$m around a Be star was 48 Lib \citep[B3 IVe;][]{Griffith2015}, a shell star \citep[i.e., viewed equator-on;][]{Hanuschik1996}. Interferometric observations constrained the orientation in space (i.e., the tilt angle with respect to the plane of the sky) of this circumstellar decretion disk \citep{Pott2010}, while its position angle was determined by linear polarimetry \citep{Stefl2012a}. Despite a difference in size of 5 orders of magnitude, the IR nebula visible in WISE seems to be parallel \citep{Griffith2015} to the circumstellar decretion disk, indicating a common axis of rotation.

Recently, an elliptical, that is, centered nebula around the classical Be star HD\,303056 was discovered in MIPS data at 24\,$\mu$m by Vasilii Gvaramadze (priv. comm.), which we will investigate in detail in a follow-up paper (Bodensteiner et al., in prep.). The presence and characteristics of such circumstellar nebulae around Be stars could give further indications pointing toward their binary origin.
   
 \subsubsection{BeXRBs and Be+sdO systems}
No aligned bow shocks were found around BeXRBs and Be+sdO systems. As discussed above, aligned bow shocks are formed by high-velocity stars, whereas the standard process of formation of BeXRBs and Be+sdO systems, mass transfer during RLOF, does not impart great velocity kicks to the systems. BeXRBs and Be+sdO systems are thus still bound in the original binary system. In the case of BeXRBs, this indicates that the binary system was not disrupted in the SN explosion of the companion. 

The centered nebulae associated with one BeXRB and two Be+sdO systems might be the left-over material lost in previous binary interactions in which the primary star evolved into a compact object and the secondary was spun up to become a Be star. The low numbers, however, prevent us from any further statistical inferences. 

\section{Conclusions}
We discovered a large number of IR nebulae associated with early-type stars. In addition to 4 BeXRBs, 3 Be+sdO systems, and 34 Be stars (with mostly unknown binary status) with extended IR nebulae, we detected 255 nebulae around OBA stars in the Bright Star Catalogue. 

Among the 255 nebulae around BSC stars, we find that 94 nebulae have the typical sickle-like shape of a bow shock that forms when the relative velocity between star and surrounding ISM is supersonic. They can be subdivided into two groups: the first group, accounting for 13\% of all nebulae, are bow shocks formed by high-velocity stars that probably gained their high velocity through close binary interactions. The second group (24\%) supports the recent development in the understanding of bow shock formation. The missing alignment between proper motion and position angle of the apex of the bow shock indicates that not only a high stellar velocity, but also large-scale motions of the ISM produce bow shocks. This leads us to the assumption that in stars with aligned bow shocks, we see the combined effect of strong wind and high space velocity, while in not-aligned bow shocks, the high space velocity of the star itself is less relevant than the turbulent ISM motion. This is in agreement with the findings by \citet{Kobulnicky2016} and demonstrates that unaligned bow shocks can be used as tracers for large-scale ISM motions.

Using aligned bow shocks as indicators for high-velocity stars, we find a high-velocity star fraction of almost 20\% in our limited sample of O stars. This agrees with earlier findings of \citet{Blaauw1961}, \citet{Gies1986}, and \citet{Tetzlaff2010}, for instance. 

Furthermore, 52 centered nebulae have round, elliptical, or diffuse shapes. While $\sim 13\%$ might be bow shocks viewed head- or tail-on, this explanation does not hold for the majority. We can exclude that the IR emission arises in the remains of the parental disk, and also that the nebulae are remnants of the SN explosion of a former companion star. Because of their low mass-loss rates, single MS B and A stars do not form dust. MS O stars and B and A supergiants can have a higher mass loss. They account only for a small fraction of the whole sample, however.

Finally, we propose that the remaining centered nebulae, that
is, 16\% of all nebulae, are formed in the context of previous binary interactions. The observed IR nebulae might be explained through dust formed during non-conservative mass transfer (either RLOF or a stellar merger) that is heated by the OBA star (and a compact companion, if merging was avoided). Such nebulae should show elevated metal abundances due to their formation history and have a short lifetime because of their expansion.

In total, we can conclude that (16+13)\%, that is, about one-quarter to one-third, of the nebulae detected around OBA stars in the BSC can be placed in the context of binary evolution. On the one hand, they can be seen as an indication of the velocity gain by binary interaction leading to the formation of a bow shock. On the other hand, the nebulae can be direct observational evidence of the material lost in non-conservative mass transfer (i.e., in RLOF or a stellar merger) heated by the star (and a compact companion). Applying this to the number of O, A, and B stars in the BSC, we find that 8\% of all O stars in the BSC, 4\% of all B stars in the BSC, and 0.1\% of all the A stars in the BSC are associated with IR nebulae that can be place in the context of binary evolution. A majority of the A stars is, however, not single or not on the MS. The small difference between the fraction of B (4\%) and Be stars (6\%) with probably binarity-related companion IR nebulae is not significant enough to postulate a dominance of binarity in the formation of Be stars.  

Our results encourage us to use the associations of IR nebulae with massive MS stars as a general tracer of past binary interactions. This could be further exploited by spectroscopic searches for companion stars. Companion detections would not only corroborate the conclusions of this paper, but would have the potential to give important new insights into the physical processes taking place in the different so far still poorly understood binary evolution and interaction channels.

\begin{acknowledgements}
We are particularly grateful to Vasilii Gvaramadze in pointing out the mid-IR nebula around HD 303056, which initiated this study, and to Thomas Rivinius and Alex C. Carciofi for helpful discussions during early stages of the project.
We would like to thank Hugues Sana and Cole C. Johnston for valuable discussions concerning the paper and the referee for very helpful comments and suggestions.
J.B. acknowledges support from the FWO\_Odysseus program under project G0F8H6N.
This publication makes use of data products from the Wide-field Infrared Survey Explorer, which is a joint project of the University of California, Los Angeles, and the Jet Propulsion Laboratory/California Institute of Technology, funded by the National Aeronautics and Space Administration.
This work is partly based on archival data obtained with the Spitzer Space Telescope, which is operated by the Jet Propulsion Laboratory, California Institute of Technology under a contract with NASA. Support for this work was provided by an award issued by JPL/Caltech. 
This research has made use of the SIMBAD database, operated at CDS, Strasbourg, France and of NASA's Astrophysics Data System Bibliographic Services.
\end{acknowledgements}

\bibliographystyle{aa} 
\bibliography{papers} 

\begin{appendix}

\section{Physical extent of bow shocks and centered nebulae}
We compare the apex distances of bow shocks in pc with the radii of centered nebulae in pc calculated from their physical size (see Tab. \ref{Tab:meas_table}), assuming that all centered nebulae are circular. We carry out a Kuiper test in order to investigate whether the underlying distributions are similar or significantly different from each other. We find that the probability that the two distributions are similar is 5\%, that is, 2$\sigma$.
   \begin{figure}\centering
   \includegraphics[width=0.99\hsize]{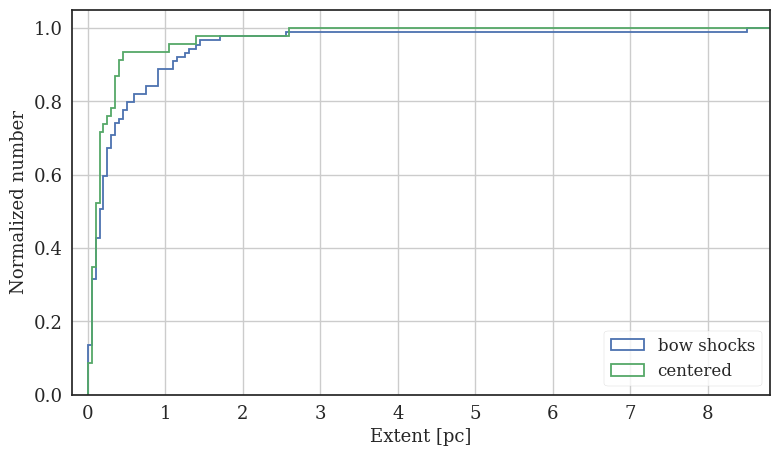}
      \caption{Normalized cumulative distribution of apex distances of bow shocks (blue) and radii of centered nebulae (green, assuming that all of them are spherical symmetric).}
         \label{Fig:r0radii}
   \end{figure}

\section{Database}
\tiny

\onecolumn
\begin{landscape}

\begin{longtable}{llllrcrrrrrrrrr}
\caption{Compilation of all parameter values from the literature. In the first column we list the identifier of the star, taken from the BSC, while the second and third columns give the stellar coordinates from SIMBAD. The column 'HD' corresponds to the HD number, while the column 'V' is the apparent V-band magnitude, also taken from SIMBAD. In the column 'spectype', we provide the spectral classification listed in SIMBAD. Parallaxes $p$ with their uncertainties $\Delta$p (columns 7 and 8) and proper motions $\mu_{Ra}$ with uncertainties $\Delta \mu_{Ra}$ and $\mu_{Dec}$ with uncertainties $\Delta \mu_{Dec}$ (columns 9 to 12) are taken from the Hipparcos catalogue \citep{VanLeeuwen2007} if available. Proper motions were converted to the rest frame of the local ISM by correcting for the Galactic rotation following \citet{Comeron2007}. Radial velocities $RV$ and uncertainties $\Delta\,RV$ (columns 13 and 14) are mainly taken from \citet{Gontcharov2006}. For stars that are not included in this catalogue, we took radial velocities from \citet{Kharchenko2007} or \citet{Pourbaix2004}, indicated by the year in the column 'ref'.}
\label{Tab:lit_table}\\
\hline
Identifier & RA & DEC & HD & V & spectype & p & $\Delta$p& $\mu_{RA}$ & $\Delta \mu_{RA}$ & $\mu_{DEC}$ & $\Delta \mu_{DEC}$ & $RV$ & $\Delta\,RV$ & ref \\ 
       & [J2000]& [J2000]  &  & [mag] & & [mas] & [mas] & [mas\,yr$^{-1}$]& [mas\,yr$^{-1}$] & [mas\,yr$^{-1}$]& [mas\,yr$^{-1}$] & [km\,s$^{-1}$] & [km\,s$^{-1}$] & \\ \hline
\endfirsthead
\caption{continued.}\\
\hline
Identifier & RA & DEC & HD & V & spectype & p & $\Delta$p& $\mu_{RA}$ & $\Delta \mu_{RA}$ & $\mu_{DEC}$ & $\Delta \mu_{DEC}$ & $RV$ & $\Delta\,RV$ & ref\\ 
       & [J2000]& [J2000]  &  & [mag] & & [mas] & [mas] & [mas\,yr$^{-1}$]& [mas\,yr$^{-1}$] & [mas\,yr$^{-1}$]& [mas\,yr$^{-1}$] & [km\,s$^{-1}$] & [km\,s$^{-1}$] & \\ \hline
\endhead
\hline
\endfoot
HR 38 & 00 12 50.25 & +37 41 37.2 & 829 & 6.7 & B3III & 1.80 & 0.42 & 1.76 & 0.46 & -2.65 & 0.30 & -5.5 & 2.4 &  \\
HD 1976 & 00 24 15.66 & +52 01 11.7 & 1976 & 5.6 & B5IV & 3.26 & 0.63 & 13.62 & 0.43 & -3.92 & 0.33 & -15.8 & 2.3 &  \\
HD 2626 & 00 30 19.93 & +59 58 39.2 & 2626 & 5.9 & B7IV & 4.24 & 0.50 & 16.26 & 0.45 & -3.29 & 0.38 & -17.4 & 0.9 &  \\
lam Cas & 00 31 46.36 & +54 31 20.2 & 2772 & 4.8 & B8Vn & 8.64 & 0.43 & 41.20 & 0.29 & -16.54 & 0.35 & -12.2 & 1.3 & 2007 \\
kap Cas & 00 32 59.99 & +62 55 54.4 & 2905 & 4.2 & B1Ia & 0.73 & 0.17 & 3.65 & 0.17 & -2.07 & 0.16 & 0.3 & 0.8 &  \\
53 Psc & 00 36 47.31 & +15 13 54.2 & 3379 & 5.9 & B2.5IV & 3.14 & 0.30 & 0.85 & 0.29 & -13.68 & 0.23 & -8.0 & 0.9 &  \\
zet Cas & 00 36 58.28 & +53 53 48.9 & 3360 & 3.7 & B2IV & 5.50 & 0.16 & 17.38 & 0.13 & -9.86 & 0.11 & -0.2 & 0.1 &  \\
ksi Cas & 00 42 04.00 & +50 30 45.1 & 3901 & 4.8 & B2V & 2.28 & 0.51 & 12.24 & 0.38 & -6.58 & 0.32 & -10.6 & 7.4 & 2007 \\
\end{longtable}

\newpage
\begin{longtable}{lllllllrrrrrrrrrrr}
\caption{Overview over all the measured and calculated parameters. The first three columns give identifier and coordinates and are the same as in Table \ref{Tab:lit_table}. The morphological classification we assigned to each nebula (with abbreviations: bow shocks = 'bs', unaligned bow shocks = 'bsna', centered = 'cent', unresolved = 'unres', and not classified nebulae = 'nc'. Columns 5 to 7 indicate whether the star is a known Be star ('Be'), has a nebulous spectrum ('n'), or is a known binary ('bin'). This information is taken from the spectral types in the BSC and the SIMBAD classification, and has to be considered with caution. The distance $d$ and its uncertainties $\Delta d$ (columns 8, 9) are determined from the parallax. The space velocity is calculated from the radial velocity, proper motion and distance. We determined the total space velocity if the individual uncertainties in distance, proper motion, and radial velocity measurements were $\leq 30\%$. The column 'size$_{ang}$' contains the angular size of the nebulae and the column 'flux' gives the total flux at 22\,$\mu$m. The angular size is converted to physical sizes (column 'size') using the distance d. Column '$\alpha$' displays the misalignment angle $\alpha$ between the proper motion of the star and the bow shock shape (see above). It is only given for bow shocks and unaligned bow shocks. The last two columns give the angular distance R(0)$_\text{ang}$ and the physical distance R(0) between the position of the star and the apex of the bow shock. }
\label{Tab:meas_table}\\
\hline
Ident & RA & DEC & class & Be & n & bin & d &  $\Delta$d& $v_\mathrm{space}$ & $\Delta v_\mathrm{space}$ & size$_\mathrm{ang}$ & flux & size & $\Delta \mathrm{size}$ & $\alpha$ & R(0)$_\mathrm{ang}$ & R(0)\\
 & [J2000]& [J2000] & & & & & [pc] & [pc]& [km\,s$^{-1}$] & [km\,s$^{-1}$]& [$'\,^2$] & [Jy] & [pc$^2$] & [pc$^2$] & [$^{\circ}$] & [$'$] & [pc]\\ \hline
\endfirsthead
\caption{continued.}\\
\hline
Ident & RA & DEC & class & Be & n & bin & d &  $\Delta$d& $v_\mathrm{space}$ & $\Delta v_\mathrm{space}$ & size$_\mathrm{ang}$ & flux & size & $\Delta \mathrm{size}$ & $\alpha$ & R(0)$_\mathrm{ang}$ & R(0)\\
 & [J2000]& [J2000] & & & & & [pc] & [pc]& [km\,s$^{-1}$] & [km\,s$^{-1}$]& [$'\,^2$] & [Jy] & [pc$^2$] & [pc$^2$] & [$^{\circ}$] & [$'$] & [pc]\\ \hline
\endhead
\hline
\endfoot
HR 38 & 00 12 50.25 & +37 41 37.2 & unres &  &  &  & 555.56 & 129.63 & 0.55 & 10.03 & 3.79 & 0.02 & 0.014 & 0.007 &  &  &  \\
HD 1976 & 00 24 15.66 & +52 01 11.7 & bsna &  &  & x & 306.75 & 59.28 & 2.16 & 25.98 & 2.55 & 0.10 & 0.017 & 0.007 & 60 & 0.7 & 0.06 \\
HD 2626 & 00 30 19.93 & +59 58 39.2 & unres &  & x & x & 235.85 & 27.81 & 2.77 & 25.44 & 1.26 & 0.23 & 0.013 & 0.003 &  &  &  \\
lam Cas & 00 31 46.36 & +54 31 20.2 & unres &  & x & x & 115.74 & 5.76 & 3.45 & 27.26 & 0.68 & 0.17 & 0.004 & 0.001 &  &  &  \\
kap Cas & 00 32 59.99 & +62 55 54.4 & bs &  &  &  & 1369.86 & 319.01 &  & 27.27 & 6.09 &  &  &  & 20 & 3.2 & 1.28 \\
53 Psc & 00 36 47.31 & +15 13 54.2 & unres &  &  &  & 318.47 & 30.43 & 1.93 & 22.20 & 1.21 & 0.11 & 0.017 & 0.003 &  &  &  \\
zet Cas & 00 36 58.28 & +53 53 48.9 & bsna &  &  &  & 181.82 & 5.29 & 36.37 & 17.24 & 0.28 & 0.88 & 0.102 & 0.006 & 80 & 1.3 & 0.07 \\
ksi Cas & 00 42 04.00 & +50 30 45.1 & bsna &  &  & x & 438.6 & 98.11 & 67.66 & 30.80 & 4.10 & 3.10 & 1.101 & 0.493 & 25 & 1.7 & 0.22 \\
\end{longtable}
\end{landscape}
\end{appendix}
\end{document}